# Comment 'On the Connection Between Planets, Dark Matter and Cancer', by Hector Socas-Navarro (arXiv:1812.02482 [physics.med-ph])

Konstantin Zioutas[*,‡], Edward Valachovic[†,§], Marios Maroudas[*]

[*] Physics Department, University of Patras, 26504 Patras, Greece
[†] Department of Epidemiology and Biostatistics, School of Public Health,
    University at Albany, State University of New York, 12144 New York, USA

[‡] zioutas@cern.ch
[§] evalachovic@albany.edu

**Abstract:**

In arXiv:1812.02482 Socas-Navarro (SN) provided multiple confirmation of the claimed ~88 days melanoma periodicity [4] (which remarkably coincides with the orbital period of Mercury). This greatly strengthens the observation by Zioutas & Valachovic (ZV). Here we comment on the work by SN, because it objects the interpretation of the observation by ZV. Notice that SN objection is based on serious assumptions, which were explicitly excluded by ZV. Further, the conclusion made with a sub-set of data (4%) is statistically not significant to dispute ZV. On the contrary, since the same periodicity appears also in other 8 major cancer types, we consider it as a global oscillatory behaviour of cancer. At this stage, such a rather ubiquitous cancer periodicity makes any discussion of a small subset of data at least secondarily. Further, we show here that the ~88 days Melanoma periodicity is not related to solar activity. Planetary lensing of streaming low speed invisible massive particles remains the only viable explanation, as it has been introduced previously with a number of physics observations [4]. We also show that planetary lensing of low speed particles cannot be considered in isolation, because of the dominating Sun's gravity, at least for the inner planets. Interestingly, gravitational lensing / deflection favours low speed particles.

In a recent paper [1], H. Socas-Navarro (SN) has re-evaluated *part* of the work "Planetary Dependence of Melanoma" by K. Zioutas and E. Valachovic (ZV) [2], using even 8 more datasets. Here we comment on the work by SN, starting with two, in our opinion, positive aspects:

**1)** **a)** SN derives a periodicity of 87.6 days (4.17/year), confirming the value of (87.4±0.76) days as it was observed for the first time by ZV in ref. [2]. Interestingly, this periodicity appears also in all 8 major cancer categories, which have been Fourier analysed by SN. Obviously, this is a diversified confirmation, which strengthens greatly the initial observation by ZV. **b)** Figure 2 of the work by SN [1] confirms previous observation of the 11 years oscillation of melanoma [3].

**2)** SN makes an extensive introduction to dark matter and WIMPs, arriving to conclusions objecting the work by ZV, since "it is incompatible with the current WIMP paradigm" [1]. We wish to stress here that the physics part of the work by ZV is based on ref.[4]; SN has apparently overlooked this important reference, since it is clarified there already in the introduction [4]:

> "…we refer to generic dark candidate constituents as "*invisible massive matter*", in order to distinguish them from ordinary dark matter."

In addition, the words 'dark matter' and 'WIMPs' are not mentioned at all by ZV, (see ref.[2]). In other words, the conclusions made by SN are based just on dark matter and WIMPs, which are excluded by ZV (and in ref. [4] too); i.e., the objections by SN are thus based on assumptions considered as inapplicable [2,4].

**3)** Melanoma and race: SN uses throughout his work the WIMP paradigm to conclude that afro-americans cannot be selectively immune to dark matter. Firstly, we repeat that WIMPs and dark matter are out of consideration by ZV. Secondly, the webpage of the US Centers for Disease Control [5] just illustrates that no race is immune to melanoma along with people of all ages, ethnicities, and sexes, which are not always affected equally. More specifically, the melanoma appearance in afro-americans makes 4% of the total rate [5]. Thus, the conclusion by SN that afro-americans should be immune to dark matter does not apply. Because, still if we refer instead to "invisible massive dark matter" as advocated by ZV, such small rates may prevent hidden signals from rising above noise. E.g., even a 10 σ signal based on the total population will be at the ~2 σ level for the statistics available with afro-americans, i.e., no conclusion can be made presently. In fact, since the same periodicity appears also in other 8 major cancer types, such a rather ubiquitous cancer periodicity makes any discussion of a small subset of data at least secondarily.

**4)** On the other counterarguments by SN (section 3.1-3.4):

**a)** To realise the admittedly missing suitable geometrical scenario in space we suggest to study first the unnoticed ref. [4]. In fact, Figure 2 in ref.[6] and Figure 2 a) in ref.[7] illustrate with their few sample trajectories how gravitational lensing of slow speed particles occurs in the solar system, which is dominated by the Sun. Apparently, the (inner) planets have also an impact on the overall gravitational focusing performance for slow speed particles (see ref. [8]). Within the scenario of slow invisible massive particles, it is reasonable to expect a modulation of the focused streaming matter downstream at the planetary orbital period, and this fits the observation by ZV.

The precise alignment of a stream for the planetary gravitational lensing to occur, which is given by SN are strongly relaxed for slow speed particles; the Einstein ring and the deflection angle increase with decreasing velocity v as $1/v$ and $1/v^2$, respectively (see ref.[9]). This is shown with the trajectories given in ref. [6,7]. Moreover, a planetary gravitational focus, of slow speed particles, can result, ideally, to a flux enhancement by as much as a factor of $10^6$ [8], while the corresponding enhancement by the Sun can be orders of magnitude larger [9]. Therefore, planetary gravitational lensing effects cannot be seen isolated from the Sun.

**b)** Concerning the rare double planetary alignment with a stream as it is arisen by SN, it applies to fast streaming matter (speeds above ~0.01c). However, for slow *invisible massive particles* (with speed below ~300 - 1000 km/s) the planetary gravitational impact must be seen in connection with the dominating gravitational force by the Sun. The aforementioned considerations (4.a)) along with both Figures 2 in ref. [6,7] illustrate the actual situation.

Moreover, planetary correlations have been observed with the dynamic Earth atmosphere [4], i.e., its degree of ionization. This is an independent signature for planetary impact at Earth's site.

**c)** Diagnosis delay and periodicity: ZV have addressed the latency issue at the end of the abstract and the main text of ref.[2]. We note that only a perfectly flat random delay distribution of more than 3 months, between the onset and the diagnosis of melanoma, could suppress the appearance of the observed short periodicity. After all its amplitude implies a small fraction (~few %) of melanoma with short latency, which is reasonable to exist. Therefore, ZV concluded that the observed 87.4 days periodicity points *in its own right* at a short latency period. Interestingly, the (multiple) confirmation of this periodicity by SN is very encouraging and strengthens the perspectives of this new approach in medicine. Investigations from the south hemisphere as pointed out by SN are of course interesting as well as the search for possible latitudinal dependence, which can be important for the identification of the assumed invisible streams.

To the best of our knowledge there is no other interpretation for the observed planetary dependence of melanoma than the driving idea of "streaming invisible massive matter" from the dark sector, which was followed-up by ZV. More importantly, this oscillatory behaviour seems ubiquitous in all Fourier analysed cancer types by SN, which is very welcome and encouraging.

ZV have focused on melanoma "only", because an 11 years periodicity had been observed before [3]. As it was demonstrated earlier [4], this points at a possible planetary correlation, which was finally found.

**5)** Solar activity: In Figure 1 (A) is shown the Fourier spectra for the melanoma monthly rate (Figure 7c in ref.[2]) along with the corresponding one for the solar line at a wavelength of 10.7 cm (F10.7) for the same time interval 1973-2010 (38 years). Note that F10.7 is widely considered as proxy for the solar activity. A comparison between both spectra (Figure 1) excludes that the observed melanoma periodicity of (87.4±0.76) days has known solar activity at its origin! Though, in Figure 1 (B) the calculated Fourier spectrum of the original daily values of F10.7 provides various other lines, but not at 88 or 365 days. This is an interesting observation. Because when we calculate the sum of consecutive 88 days and 365 days (Figure 2) of the daily intensity values of the solar line F10.7 (~2.8 GHz), both spectra show a rich structure. Though, the Fourier analysis does not show a peak at 88 or 365 days, which demonstrates the limit of the Fourier analysis in this case. At the same time, the spectral richness of spectra like both in Figure 2 show the advantage of this simplified spectral analysis. In fact, several such spectra have been included in ref.[2].

# FIGURES

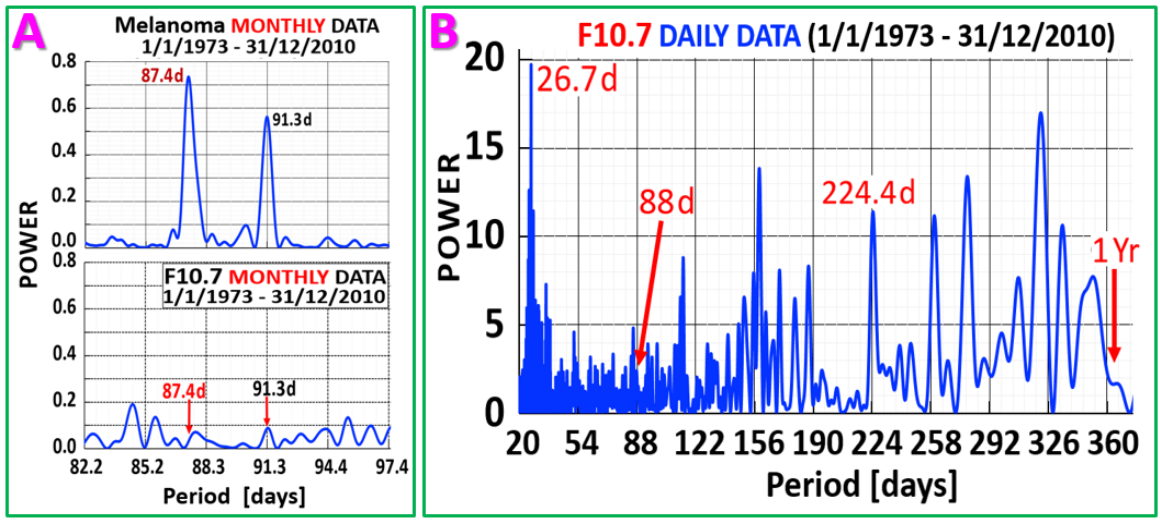

**Figure 1**   (A) The Fourier spectrum of the monthly melanoma rate around the periodicity of 80 to100 days (upper spectrum). The lower spectrum shows the Fourier spectrum of the intensity of the solar line around 2.8 GHz, or F10.7 line using the monthly values (lower spectrum).  (B) The Fourier spectrum of the original daily data of F10.7 for the same time interval. No line is seen at 88 days and 365 days (see Figure 2 for comparison).

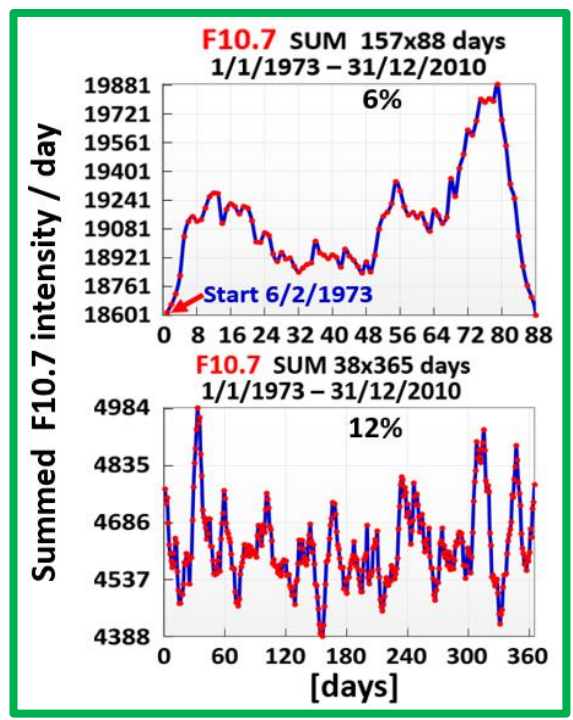

**Figure 2**   Solar activity: the sum of the daily intensity of the solar F10.7 line of consecutive time intervals of 88 days (157x88 days, upper spectrum) and 365 days (38x365 days, lower spectrum) [10].